\documentclass[11pt,twoside]{article}
\usepackage{asp2010}

\resetcounters

\begin{document}

\title{Lessons learnt in building VO resources: binding together several VO
standards into an operational service}
\author{Igor~Chilingarian$^{1,2}$, Fran\c cois~Bonnarel$^3$, 
Mireille Louys$^3$, and Pierre~Le~Sidaner$^4$
\affil{$^1$SAO, Harvard-Smithsonian Center
for Astrophysics, 60 Garden Street MS09, 02138 Cambridge, MA, USA}
\affil{$^2$Stenberg Astronomical Institute, Moscow State University,
13 Universitetsky Prospect, 119992 Moscow, Russia}
\affil{$^3$CDS, Observatoire astronomique de Strasbourg,
11 rue de l'Universit\'e, 67000 Strasbourg, France}
\affil{$^4$VO Paris Data Centre, Observatoire de Paris-Meudon, 
61 ave. de l'Observatoire, 75014 Paris, France}
}

\begin{abstract}
The International Virtual Observatory Alliance (IVOA) developed numerous
interoperability standards during the last several years. Most of them are
quite simple to implement from the technical point of view and even contain
``SIMPLE'' in the title. Does it mean that it is also simple to build a
working VO resource using those standards? Yes and no. ``Yes'' because the
standards are indeed simple, and ``no'' because usually one needs to implement
a lot more than it was thought in the beginning of the project so the time
management of the team becomes difficult. In our presentation we will start
with a basic case of a simple spectral data collection. Then we will
describe several examples of small'' technologically advanced VO resources
built in CDS and VO-Paris and will show that many standards are hidden from
managers' eyes at the initial stage of the project development. The projects
will be: (1) the GalMer database providing access to the results of
numerical simulations of galaxy interactions; (2) the full spectrum fitting
service that allows one to extract internal kinematics and stellar
populations from spectra of galaxies available in the VO. We conclude that:
(a) with the existing set of IVOA standards one can already build very
advanced VO-enabled archives and tools useful for scientists; (b) managers
have to be very careful when estimating the project development timelines
for VO-enabled resources.
\end{abstract}

\section{Introduction} 

On the side of IVOA we now have a comprehensive set of
standards\citep{AGftITCG11} including: data formats (VOTable,
\citealp{VOTable}), VO resource description (Resource Metadata), data models
for 1D spectra (Spectrum Data Model) and simulations (SimDM/SimDB) and much
more complex and general Characterisation Data Model \citep{CharDM},
Astronomical Data Query Language, protocols to access tabular data (TAP,
\citealp{TAP}), images (SIAP), and spectra (SSAP, \citealp{SSAP}), a Simple
Application Messaging Protocol \citep{SAMP} that allows different VO tools
and services to talk to each other, authorisation and authentication
mechanisms, and others. Some other standards are still at different phases
of development. Now it became possible to handle even very complex
astronomical datasets in the Virtual Observatory, such as 3D spectroscopy
\citep{CBLM06,ChilVO08} and results of N-body simulations. Some of the
standards even carry the adjective ``simple'' in their titles.

Is it really simple and fast to implement operational VO services using
these standards? Unfortunately, many (mostly technical) aspects are often
hidden from the view. And the implementation of these details may take
significant amount of time.

\section{Examples}


\subsection{Spectral Data Collection}

\begin{figure}
\includegraphics[width=0.73\hsize]{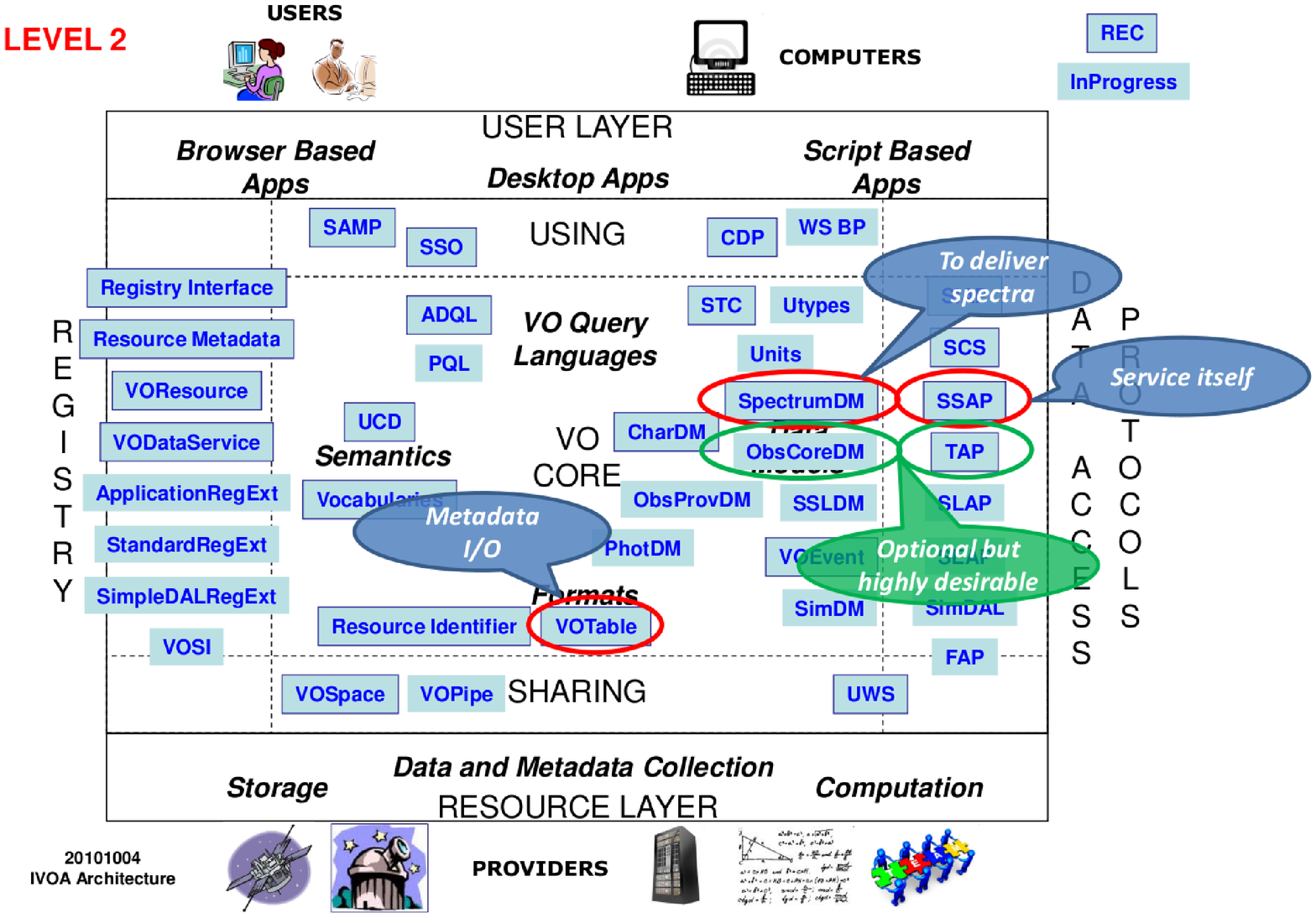}
\caption{IVOA standards required (red) and recommended (green) to implement a simple spectral data
collection access by a data centre.\label{figsdc}}
\end{figure}

The simplest case is a collection of 1-dimensional spectral data. One may
think that such a service requires the implementation only of the SSAP.
However, to make I/O of the spectral data one needs to implement the
Spectrum Data Model and its serializations including VOTable. At present,
the implementation of the TAP access using the Observation DM Core
Components \citep{ObsCore} is highly desirable. All standards that one needs to
implement in order to build an operational VO spectrum data access service
are shown in Fig.~\ref{figsdc} on the VO Infrastructure Diagram.

\subsection{The GalMer Database}

The first ``technologically advanced'' project that we describe here is the
GalMer database. The GalMer database is a part of the Horizon project,
providing access to a library of TreeSPH simulations of galaxy interactions
\cite{Chilingarian+10}. We have developed a set of value-added tools related
to data visualization and post-processing with available VO-interfaces,
including the spectrophotometric modelling of galaxy properties, making
GalMer the most advanced resource providing online access to the results of
numerical simulations. These tools allow direct comparison of simulations
with imaging and spectroscopic observations (see e.g.
\citealp{Chilingarian+09b}).

\begin{figure}
\includegraphics[width=0.73\hsize]{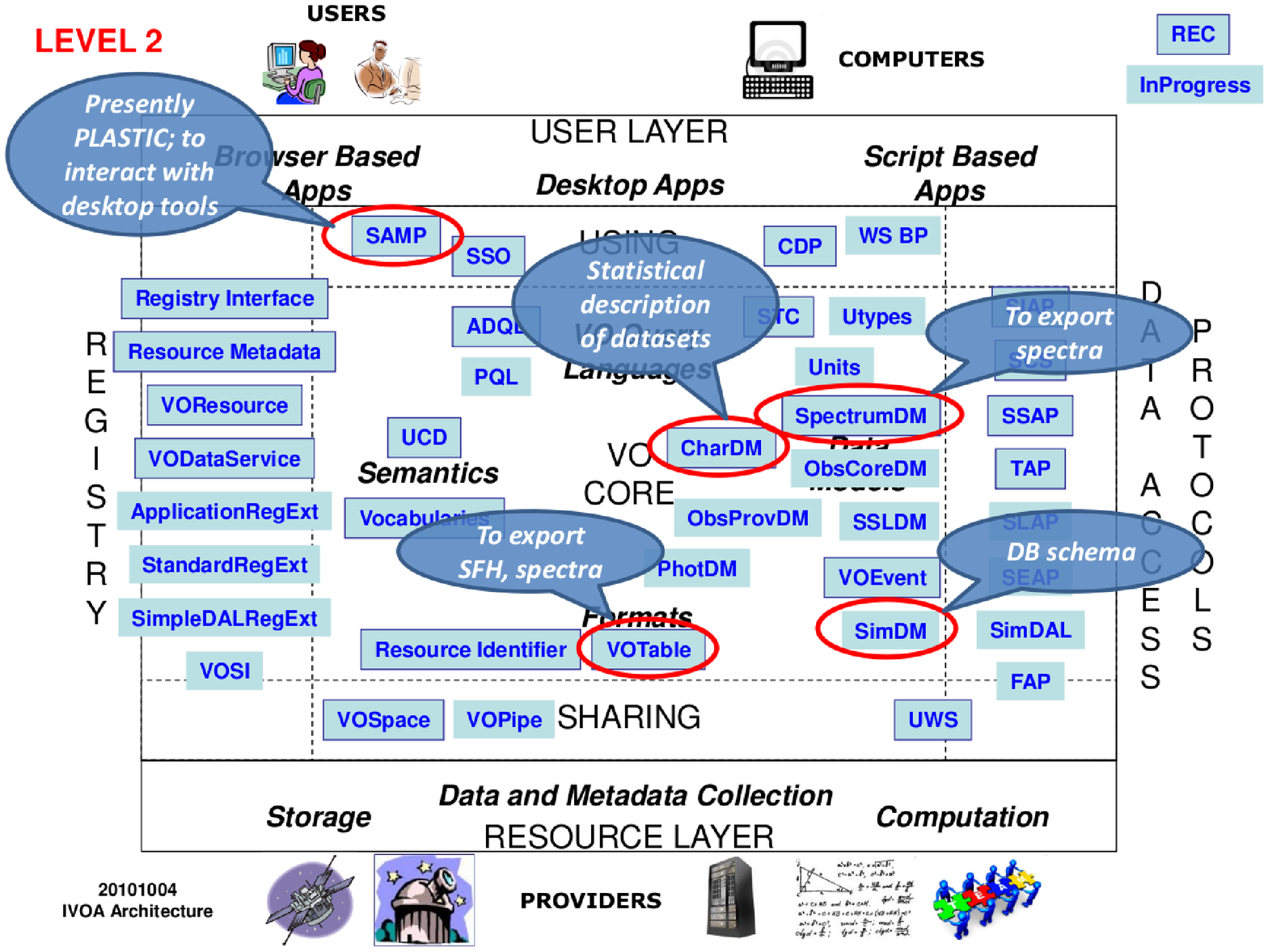}
\caption{IVOA standards required to implement the GalMer database and
services.\label{figgalmer}}
\end{figure}

The database schema of GalMer is based on the SimDM data model (see
Fig.~\ref{figgalmer}). We use Characterisation DM in order to provide the
advanced statistical description of datasets. Spectrum Data Model is used
for the ``virtual telescope'' service, i.e. the on-the-fly
spectrophotometric modelling. Spectral and imaging data are displayed using
an integration of a browser with desktop VO tools, we use PLASTIC, a SAMP
precursor.

\subsection{Full Spectrum Fitting of 1D Spectra of Galaxies}

The last example is an operational VO service providing access to the full
spectrum fitting service implementing the ``Penalized Pixel Fitting''
technique \citep{CE04}. This tool allows one to fit a set of stellar
population models against an observed 1D absorption line spectrum of a
galaxy and determine its internal kinematics (radial velocity, velocity
dispersion and higher order moment of the line-of-sight velocity
distribution, see \citealp{vdMF93}). We are presently working on the
implementation of the NBursts pixel fitting algorithm \citep{CPSA07,CPSK07}
that also returns stellar population parameters in the same minimization
loop.

\begin{figure}
\includegraphics[width=0.73\hsize]{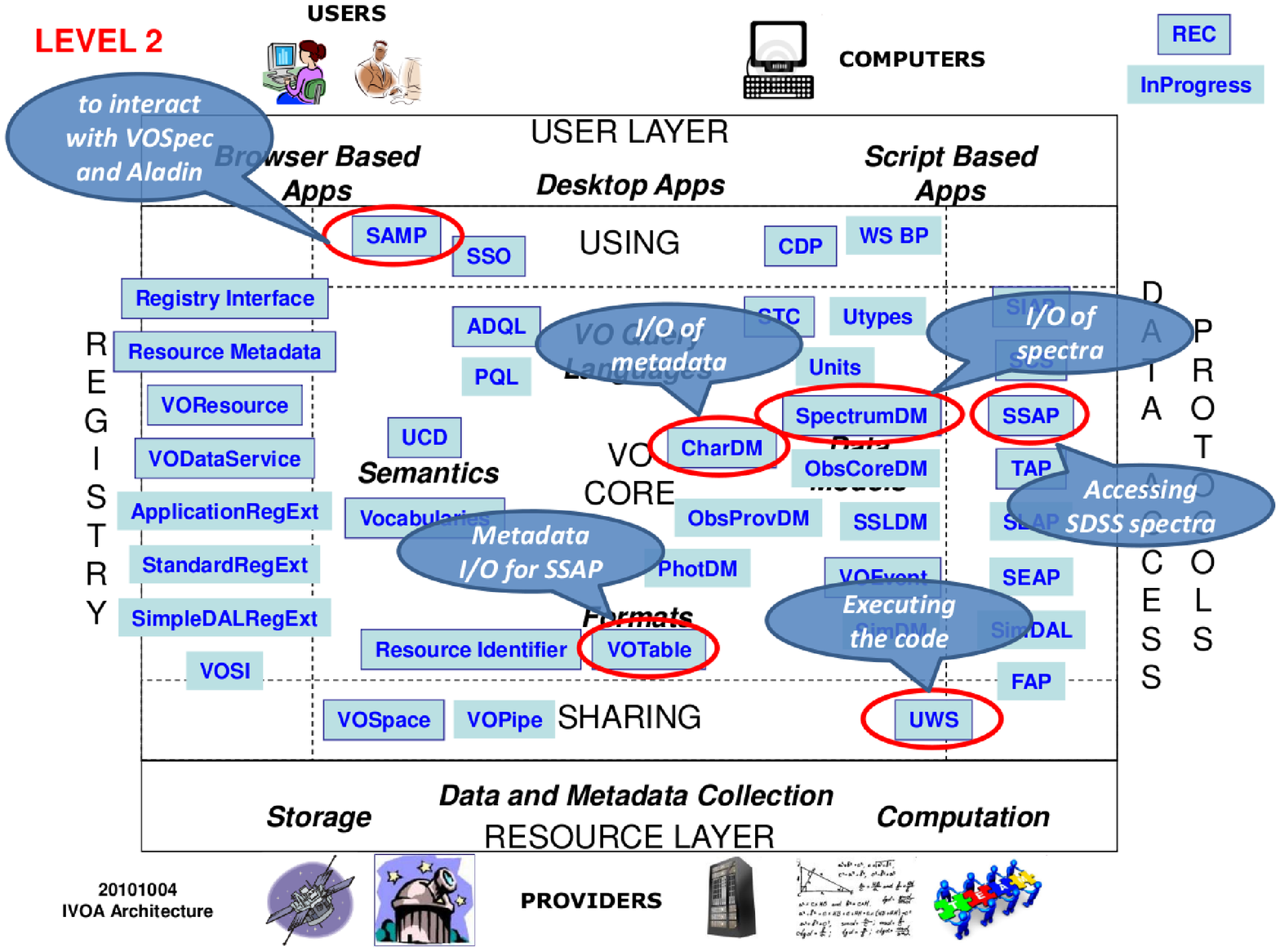}
\caption{IVOA standards required to implement the full
spectrum fitting service.\label{figfitting}}
\end{figure}

We implemented our service for the Sloan Digital Sky Survey \citep{SDSS_DR7}
data (see Fig.~\ref{figfitting}) as a Universal Worker Service \citep{UWS} that
executes a compiled binary on a server. In addition to UWS, we had to
implement SSAP over SDSS data, and SAMP to interact with Aladin and VOSpec.

\section{Summary}

Our conclusion is that in many cases with the existing set of IVOA standards one
can already build very advanced VO-enabled archives and tools useful for
scientists. However, project managers have to be very careful when
estimating the project development timelines for VO-enabled resources as a
significant manpower overhead may be needed.


\bibliography{P023}

\end{document}